\documentclass[aps,pre,floatfix,twocolumn,nofootinbib]{revtex4}
\usepackage{graphicx} 
\usepackage{amsmath}\usepackage{amssymb}
\usepackage{subfigure}
\usepackage{epstopdf}

\newcommand{\dd}{{\rm d}}
\newcommand{\comment}[1]{}
\newcommand{\be}{\begin{equation}}
\newcommand{\ee}{\end{equation}}

\newcommand{\bean}{\begin{eqnarray}}          
\newcommand{\eean}{\end{eqnarray}}

\newcommand{\bea}{\begin{eqnarray*}}          
\newcommand{\eea}{\end{eqnarray*}}

\begin{document}

\title{Time arrow is influenced by the dark energy}

\author{A.E. Allahverdyan$^{1}$ and
  V.G. Gurzadyan$^{1,2,3}$ 
}

\address{$^{1}$Department of Theoretical Physics and Center for Cosmology and Astrophysics, Yerevan Physics Institute, Yerevan 0036, Armenia\\
$^{2}$Center for Cosmology and Astrophysics, Yerevan State University, Yerevan 0025, Armenia\\
$^{3}$SIA, Sapienza University of Rome, Rome I-00185, Italy}

\begin{abstract} The arrow of time and the accelerated expansion are
  two fundamental empirical facts of the Universe.  We advance the
  viewpoint that the dark energy (positive cosmological constant)
  accelerating the expansion of the Universe also supports the time
  asymmetry. It is related to the decay of meta-stable states under
  generic perturbations, as we show on example of a microcanonical
  ensemble. These states will not be meta-stable without dark
  energy. The latter also ensures a hyperbolic motion leading to
  dynamic entropy production with the rate determined by the
  cosmological constant.

\end{abstract}


\maketitle

\section{Introduction}

Astronomical data point out the existence of a positive cosmological
constant $\Lambda$ (dark energy).
 
The precise origin of $\Lambda$ is not yet clear. Various models attribute it to
macroscopic vacuum fluid, geometrical term, scalar fields or modified
gravity \cite{review}. An influential scenario for $\Lambda>0$ is that 
it emerges due to vacuum fluctuations
which are able to induce negative pressure \cite{Zeld}; see e.g.
\cite{GX,DG} for the numerical fit of estimations and the observed
density of dark energy.  The positive $\Lambda$ is among the
necessary conditions along with the 2nd law also in the Conformal 
cyclic cosmology  \cite{Pen,GP}.

The time is ripe for asking about the implications of $\Lambda>0$
in basic physics. We aim to show that $\Lambda>0$, in contrast to
$\Lambda\leq 0$, leads to a specific mechanism for an emergent
thermodynamic arrow of time and entropy generation. It is related to
parametric instability of the bound, gravitating motion that becomes
metastable with respect to generic perturbations in the presence of
dark energy.

The time asymmetry and the 2nd law of thermodynamics are among the
basic empirical facts on the Nature. It is known that laws of physics
are invariant with respect to time-inversion
\cite{P,Z,Hall,gold,gal-or,lindblad,balian,ellis,earman} (more precisely they
are only CPT-invariant; the difference between T and CPT is an
important one \cite{sharp}, but it is relevant only for high energies
\cite{tdlee}). However, there are several fundamental types of motion
whose time-inversion is not observed. They are called arrows of time
\cite{P,Z,Hall,gold,gal-or,lindblad,balian,ellis,earman}:

-- thermodynamical: increase of entropy in a closed evolving
system;

-- electrodynamical: physics is dominated by retarded potentials,
although advanced potentials are formally allowed, they are not
observed;

-- cosmological: expansion of the Universe;

-- quantum-mechanical: the apearance of definite measurement results
accompanied by reduction of quantum state. If quantum mechanics is an
emergent theory, this arrow may be caused by a sub-quantum one; see
\cite{subq_1,subq_2} for possible scenarios.


There are certain relations between the arrows \cite{gold}; in
particular, the quantum-mechanical arrow can to an extent be reduced
to the thermodynamic one \cite{abn,ellis}. Recent research clarified
the place of this arrow in microscopic dynamics \cite{parrondo,feng}
and its relation with external perturbations \cite{mahler}.

In all arrows there are two closely related aspects: initial
conditions and the proper dynamics. {Let us recall and
  illustrate this point via the emergence of the thermodynamic arrow
  within the system-bath approach \cite{AG} from the T-invariant
  Hamiltonian dynamics; see \cite{gal-or,lindblad,balian,ellis} for a
  general background.} It was argued in \cite{AG} that {\it (i)} the
thermodynamical time arrow in the system can arise in the system due
to the limited observability of the bath; {\it (ii)} while the initial
conditions are necessary for the emerging of a {\it pre-arrow}, the
full time arrow will be established if also the dynamics of the system
is Markovian (no-memory). Namely, when a quantum system ${\rm S}$
interacts with a thermal bath ${\rm B}$, the total Hamiltonian is
split $H=H_{\rm S}+H_{\rm B}+H_{\rm I}$, between, respectively, ${\rm
  S}$, ${\rm B}$ and interaction. The state of ${\rm S}+{\rm B}$ is
described by the density matrix ${\cal D}(t)$ and the von Neumann
equation
$$
i\hbar\partial _t{\cal D}(t) =H{\cal D}(t)-{\cal D}(t)H.
$$
When the initial state at $t=0$ can be split as 
$${\cal D}(0)=D_{\rm S}(0)\otimes D_{\rm B}(0),$$ 
and the state of the system at arbitrary time $t$ is given by partial
density matrix $D_{\rm S}(t)={\rm tr}_{\rm B}{\cal D}(t)$, where ${\rm
  tr}_{\rm B}$ is the trace over the Hilbert space of the bath, one
can see the emergence of the thermodynamical arrow in the system due
to the bath's incomplete observability.

In \cite{AG} we also discussed the hyperbolicity as a possible
mechanism for the Markovian dynamics. One scenario for this relates to
the voids|underdense regions in the Universe|that are able to induce
hyperbolicity of the null geodesics even if the global spatial
curvature is zero (i.e. in the flat Universe) \cite{GK}. The
properties of the cosmic microwave background \cite{GDB} appear to fit
the observed void structure in the large scale galaxy distribution,
including in the case of the Cold Spot as a supervoid \cite{Cold,Sz}.

We now make the next step in that approach of the emergence of the
time arrow i.e. involving one more fundamental empirical fact, the
dark energy. 

We adopt an {\it Ansatz} that the dark energy acts as a bath for the
observed Universe, thus supporting the emergence of the time
arrow. The system-bath interaction has to be small in order not to
distort the state of the system. This agrees with the empirical
situation of the dark energy when the role of its influence on typical
macro-physical processes remains unnoticed both due to the low value
of its energy density and the still ambiguous cross-section of
interaction with elementary particles. In this sense our laboratory
physics reveals itself within intermediate scales on which both the
cosmological constant and vacuum fluctuations are not easily noticed,
although possibly being themselves mutually linked.

Here the relation between the arrow of time and dark energy
($\Lambda>0$) will be established for a particular scheme, though
various considerations of this {\it Anzats} can be possible. Within
this scheme, the dark energy facilitates the thermodynamic arrow of
time on those scales, where its influence can be comparable to
gravity.

\section{ Set-up}  

We consider the limit of the Newtonian gravity, where
$N$ non-relativistic test particles move in a field of a large mass $M$.
Here $\Lambda$ reveals itself as an additional parabolic potential
imposed on the usual inverse-square-law interaction
\cite{petrosian,carrera,gibbons,vg_1985}. As usual in equilibrium
statistical mechanics, we assume that initially the $N$-particle
system performs a finite motion and can be described by a
microcanonical, equilibrium distribution at some fixed energy
\cite{berdi,ll}. So no arrow of time is present initially. Our aim is
to look for two scenarios of perturbing this system such that for
$\Lambda\leq 0$ (negative or zero cosmological constant) the system is
stable and continues to perform a finite motion. These scenarios
amount, respectively, to fast and slow perturbations.  Moreover, for
the slow perturbation scenario $\Lambda\leq 0$ implies that the system
returns to exactly the same state as before the perturbation. However,
for both considered perturbation scenarios, $\Lambda>0$ (positive
cosmological constant as observed in our universe) leads to changing
the finite motion to an infinite one so that the ergodivity is
violated, i.e. the system moves in one ergodic component, and the real
motion is not anymore similar to the time-inverted one. Consequently,
the dynamic entropy increases with the rate $\propto \sqrt{\Lambda}$.

Since all these effects relate to $\Lambda$, we shall choose to work
with the simplest situation $N=1$: one-particle system for which the
initial microcanonical distribution is well-defined. All the obtained
effects exist also for $N>1$.

Thus for a test mass $m$ in the field of a larger mass $M\gg m$ the
Newton equations read: 
\begin{eqnarray}
  \label{eq:9}
\ddot{R}=-\partial \Phi(R)/\partial R,   
\end{eqnarray}
where
$R$ is the interparticle distance, and
\cite{petrosian,carrera,gibbons,bacry,vg_1985}
\begin{eqnarray}
  \label{eq:1}
\Phi (R)= 
\frac{L^2}{2m^2R^2}
-\frac{GM}{R}
-\frac{4\pi G\rho_V}{3}\,R^2.
\end{eqnarray}
Here $m\Phi (R)$ is the potential energy of the test particle, $L$ is
the (constant) orbital momentum, $-\frac{GM}{R}$ is the gravitational
attraction, and $-\frac{4\pi G\rho_V}{3}\,R^2$ is the potential
generated by cosmological term. It is characterized by 
\begin{eqnarray}
  \label{eq:10}
  \rho_V=\Lambda
  c^2/(8\pi G),  
\end{eqnarray}
the mass density of the vacuum fluid, if $\Lambda>0$ (dark energy) is
interpreted in this way. Note that in
$-\frac{G}{R}(M+\frac{4\pi\rho_V}{3}\,R^3)$, the contribution from the
dark energy is seen to arise from a homogeneous distribution with
density $\rho_V$.

The recent estimate for the dark energy density by the Planck's data
yields $0.69$ fraction of the total density \cite{Planck}.

Thus the conserved energy of the test particle is a
\begin{eqnarray}
  \label{eq:0}
E=\frac{\dot{R}^2}{2}+  \Phi (R),
\end{eqnarray}
The terms $-\frac{GM}{R}$ and $-\frac{4\pi G\rho_V}{3}\,R^2$ in
(\ref{eq:1}) are similar to each other \cite{vg_1985}: they both hold
the Newton's shell theorem (they are the only potentials having this
feature) and they possess an additional symmetry leading to closed
orbits. 

Let us mention somewhat different interpretation of (\ref{eq:1},
\ref{eq:0}): they apply to a test mass in the homogeneous, isotropic
universe \cite{30}, where the motion of the test mass is influenced
only by the matter mass $M$ and the ``vacuum'' mass $\frac{4\pi
  G\rho_V}{3}\,R^2$ inside of the sphere with the radius $R$.

Also, the third term in (\ref{eq:1}) leads to the
$\Lambda$-term in Friedmann equation \cite{gibbons}:
\begin{eqnarray}
  \label{eq:00}
\dot{a}^2=\frac{1}{a}(const - \kappa a + \frac{\Lambda}{3}),
\end{eqnarray}
where $a$ is the scale factor and $\kappa$ is a constant.  This
approach to the large scale limit for the Newtonian potential removes
the infinities peculiar to the purely Newtonian cosmology
\cite{gibbons}. Its radial dependence can become a subject of
dedicated astronomical testing based on the dynamics of galactic
halos, galaxy groups and clusters. 

This is related also to the already discussed scale (e.g. \cite{bala}),
when N-body effects become comparable to the dark energy one. The
possibility of observing the dark energy at this scale was recently
discussed in \cite{chernin_2013}.

\section{ Perturbations}

We turn to a detailed investigation of (\ref{eq:1}). First of all,
note that the term $\frac{L^2}{2m^2R^2}$ in (\ref{eq:1}) is needed for
ensuring a bounded motion of the test particle: otherwise, for $L=0$ it will
simply fall into the central mass. Apart of that, the term does not play
any important role in our study and for simplicity we replace it by a
hard-wall imposed at relatively small distance $R_0$. 

\begin{figure}[htb]
    \includegraphics[width=0.8\columnwidth]{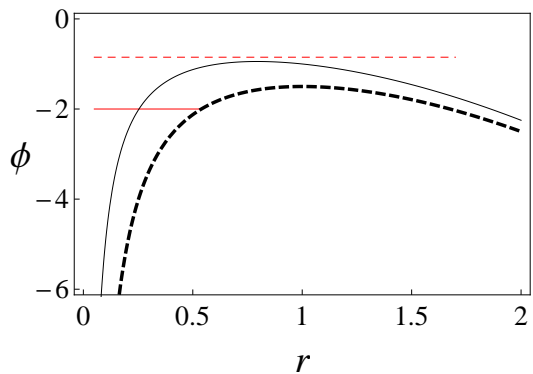} 
    \caption{The effective potential $\phi(r)$ versus distance $r$ for
      $r_0=0.05$. Dashed black curve: $\alpha=\beta=1$. Black curve:
      $\alpha=0.5$, $\beta=1$. The energy $\epsilon=-2$ refers to a
      bound motion for $\alpha=\beta=1$, but when slowly decreasing
      $\alpha(t)$ this motion becomes unbounded, i.e. its energy
      raises above the red-dashed (upper horizontal) line;
      cf. Fig.~\ref{f1c}.   }
     \label{f1b}
\end{figure}

Let us introduce characteristic scales $\bar{E}$ and $\bar{R}$ for the
energy and distance respectively, and write $\Phi(R)=\bar{E}\phi(r)$
in terms of dimensionless $r=R/\bar{R}$ and $\phi(r)$:
\begin{eqnarray}
  \label{eq:2}
  \phi(r)=-\alpha/r-\beta r^2/2,~~ r\geq r_0. 
\end{eqnarray}
where $\alpha\propto M$, $\beta\propto \rho_V$, and $r\geq r_0$ is the
hard-wall condition. Fig.~\ref{f1b} shows the form of this
potential. It is maximal at
\begin{eqnarray}
  \label{eq:7}
r_c=\alpha^{1/3}\beta^{-1/3}, ~~
\phi(r_c)=-\frac{3}{2}\alpha^{2/3}\beta^{1/3}.   
\end{eqnarray}
Hence the energies $\phi(r_c)>\epsilon>\phi(r_0)$
[$\phi(r_c)<\epsilon$] refer to bounded [unbounded] motion.

How the situation changes when the mass $M$ is time-dependent?  This
is the only natural parametric perturbation for this problem.  Indeed,
if several large masses are there (and the test particle moves in the
effective field generated by them) the inverted harmonic potential
acting on the test particle should originate from the inertia center
of the overall system $\propto -R^2_i$ \cite{bacry}, and hence it
cannot become (parametrically) time-dependent (the inertia center is
at rest). We stress that the time-dependent mass $M(t)$
  always stays much larger than the mass of the test particle.

Now provided that the test particle's motion is bounded and $M(t)$ is
slow, $E(t)$ can be described via the (adiabatic) invariant
phase-space volume \cite{hertz,ll}
\begin{eqnarray}
  \label{eq:3}
  \int \dd R\,\dd P\, \vartheta\left[E(t)-\frac{P^2}{2}-\Phi(R,M(t))\right],
\end{eqnarray}
where $P$ is the momentum [cf. (\ref{eq:0})], and $\vartheta[x]=0$ for
$x\leq 0$ and $\vartheta[x]=1$ for $x>0$ is the step-function. The
conservation of (\ref{eq:3}) relates to the fact that (in the present
case) the bounded motion is ergodic \cite{an}. Recall that the entropy
of a microcanonical equilibrium state is defined via the logarithm of
(\ref{eq:3}) \cite{berdi}. Its adiabatic conservation relates to the
second law \cite{an}. Eq.~(\ref{eq:3}) and its generalizations appear
as well in the control theory \cite{as}.

Integrating over $P$ in (\ref{eq:3}) and going to the dimensionless
quantities we get from (\ref{eq:2}) that [up to irrelevant constants]
(\ref{eq:3}) reduces to
\begin{eqnarray}
  \label{eq:4}
  J=\int_{r_0}^{\hat{r}(t)} \dd r\, \sqrt{\epsilon-\phi(r,\alpha(t))},
\end{eqnarray}
where $\epsilon(t)=E(t)/\bar{E}$, and $\hat{r}(t)$ is the maximal
distance for the finite motion at time $t$:
\begin{eqnarray}
  \label{eq:6}
\epsilon(t)=\phi(\hat{r}(t),\alpha(t)).
\end{eqnarray}
Note that $\hat{r}(t)$ is always smaller than the largest possible
distance of the bounded motion for a given $\alpha(t)$ and $\beta(t)$:
\begin{eqnarray}
  \label{eq:11}
  \hat{r}(t)\leq r_c(t)=\alpha^{1/3}(t)\beta^{-1/3}(t).   
\end{eqnarray}

Thus when changing $\alpha$ from $\alpha_1$ to $\alpha_2$, the final
energy $\epsilon_2$ is to be determined from
\begin{eqnarray}
  \label{eq:5}
  J(\alpha_1, \epsilon_1)=  J(\alpha_2, \epsilon_2).
\end{eqnarray}

A time-dependent $\alpha(t)\propto M(t)$ leads to the following two
scenarios by which the bounded motion can turn to unbounded one. Both
of them do not exist for $\Lambda\propto \rho_V\leq 0$.

\begin{figure}[htb]
    \includegraphics[width=0.8\columnwidth]{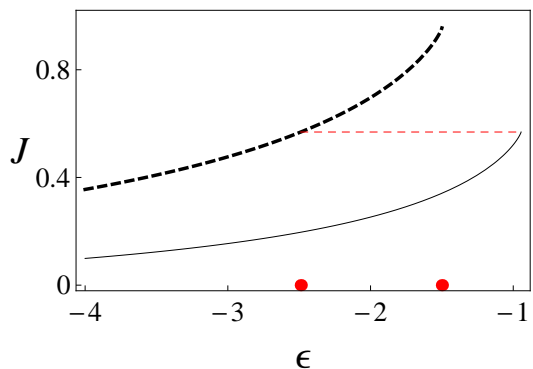} 
    \caption{The adiabatic invariant $J(\epsilon)$ as a function of
      the dimensionless energy $\epsilon$ for $\alpha=\beta=1$ (dashed
      black curve) and for $\alpha=0.5$, $\beta=1$ (black curve). For
      both curves $\phi(r_c)>\epsilon>\phi(r_0)$, where $\phi(r_c)$
      ($\phi(r_0)$) is the largest (smallest) energy for the bounded
      motion. Now energies $\epsilon\in (-1.5,-2.4889)$ (this interval
      is denoted by two red, bold points) that refer to finite motion
      under $\alpha=\beta=1$ correspond to infinite motion when
      $\alpha$ slowly decreases from $\alpha=1$ to $\alpha=0.5$. The
      range $\epsilon\in (-1.5,-2.4889)$ is additionally illustrated
      by the horizontal (dashed red) line; cf. Fig.~\ref{f1b}.  }
     \label{f1c}
\end{figure}

\begin{figure}[htb]
     \includegraphics[width=0.8\columnwidth]{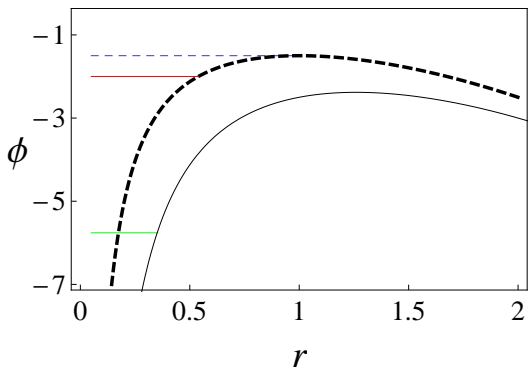} 
     \caption{The effective potential $\phi(r)$ versus distance $r$
       for $r_0=0.05$ [cf.~(\ref{eq:2})]. Dashed black curve:
       $\alpha=\beta=1$. Blue dashed line refers to
       $\phi_c=-\frac{3}{2}\alpha^{2/3}\beta^{1/3}$ (for
       $\alpha=\beta=1$); cf.~(\ref{eq:7}). For all energies below
       (above) $\phi_c$ the motion is bound (unbound). Red line: an
       example of finite motion at energy $\epsilon=-2$.  Black curve:
       $\phi(r)$ for $\alpha=2$, $\beta=1$. When $\alpha$ slowly
       changes from $\alpha=1$ to $\alpha=2$, the energy (given
       initially by the red horizontal line) decreases and always
       refers to finite motion; for $\alpha=2$ it is given by the
       green (lower horizontal) line and is equal to $-5.7593$, as
       found from (\ref{eq:5}). But if $\alpha$ changes sufficiently
       fast, the initial energy does not change much \cite{an} and now
       it corresponds to an unbound motion; compare red dashed line
       with the black curve.}
\label{f1a}
\end{figure}

\begin{figure}[htb]
\includegraphics[width=0.8\columnwidth]{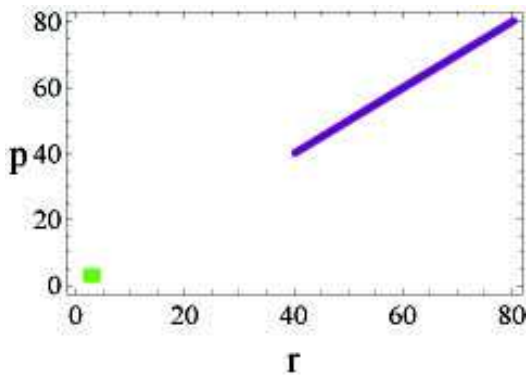} 
\caption{Phase-space density for the inverse parabolic potential with
  Hamiltonian $H(r,p)=\frac{1}{2}(p^2-r^2)$, where $(r,p)$ are
  canonically conjugate coordinate and momentum;
  cf.~(\ref{eq:8}). Green square $[(2,4); (2,4)]$ is the phase-space
  volume at the initial time $t=0$. Purple figure is the phase-space
  volume at time $t=3$. The initial volume was filled by $10^5$ random
  points; each point was given a small but finite width, defining a
  coarse-graining procedure. The evolution of each point was then
  followed from $t=0$ till $t=3$. It is seen that due to the
  coarse-graining the phase-space volume increases. }
\label{f2}
\end{figure}

{\bf 1.} Let $M(t)\propto \alpha(t)$ decrease slowly.  This can
model slow evaporation taking place from the mass $M$. It is
now possible that the bounded motion with sufficiently high initial
$\epsilon$ becomes unbounded; see Fig.~\ref{f1b}. This is related to
the fact that (\ref{eq:5}) does not have solutions $\epsilon_2$ for a
range of $\epsilon_1$ that initially were sufficiently close to
$\phi(r_c)$; see Fig.~\ref{f1c}. In other words, during the slow
decrease of $\alpha(t)$, $\epsilon(t)$ grows faster than
$\phi(r_c(t))$. Now it is crucial that the change of $\alpha(t)$ is
not very fast; otherwise $\epsilon(t)$ will not change much and will
stay bounded. Again $\rho_V>0$ is crucial for this scenario. For
$\Lambda\leq 0$, the influence of a cyclic change of $\alpha(t)$ on
the system can be made arbitrary small, provided that it is
sufficiently slow.

{\bf 2.} Let $M(t)\propto \alpha(t)$ increase. This can account for
situations with accretion driven increase of $M$. {Now $r_c$
  increases [see (\ref{eq:7})] and $\epsilon(t)$ decreases. But if
  $\alpha(t)$ changes suddenly, $\epsilon(t)$ will not change much and
  the motion will become unbounded; see Fig.~\ref{f1a}. (The energy
  $\epsilon(t)$ will not change, since during a sudden perturbation
  the force changes by a finite amount in a small time-interval. Hence
  the acceleration also changes by a finite amount, while the changes
  of the coordinate and velocity are negligible; this is the general
  feature of sudden perturbations.)

  Thus, there is a larger class of perturbations such that $\alpha(t)$
  changes not slowly, and $\epsilon(t)$ decreases slower than
  $\phi(r_c(t))$; see Fig.~\ref{f1a}.} Hence it is possible that
$\epsilon(t)>\phi(r_c(t))$ at some time $t$, and the motion becomes
unbounded. This conclusion may seem counter-intuitive: once the mass
$M$ increases, the attraction towards the center becomes stronger, but
the test particle can escape the attracting center benefiting from the
repulsion by the dark energy inside the shell.

It is crucial for this scenario that $\alpha(t)$ increases not
slowly. Otherwise, (\ref{eq:5}) predicts bounded motion; see
Fig.~\ref{f1a}. This scenario of parametric instability is due to
$\rho_V\propto \Lambda>0$, e.g. the situation with $\Lambda =0$ (no
dark energy) or with $\Lambda<0$ (negative cosmological constant) is
stable with respect to this parametric perturbation. Note that when
$\alpha(t)$ returns to its initial value|i.e. when the perturbation is
over|the motion is not turned back to bounded.

Thus in both scenarios the motion changes from bounded to
unbounded. The latter is not ergodic, e.g., only one component of the
momentum space is explored. (For example, in the 1d situation the
momentum space $P$ has two components $P>0$ and $P<0$.)  Similar
examples of irreversibility generated by non-ergodicity were analyzed
in \cite{an}.

\section{ Entropy production}

What happens with the test particle once it escapes the bounded part
of the potential? The dominant part of potential is now $-\beta
r^2/2$ and the motion generated by Hamiltonian
$H=\frac{1}{2}(p^2-\beta r^2)$, 
\begin{eqnarray}
  \label{eq:8}
\dot{r}=p,\qquad \dot{p}=\beta r,  
\end{eqnarray}
is hyperbolic: it has a positive Lyapunov exponent
$\sqrt{\beta}\propto \sqrt{\Lambda}$. Due to the second Lyapunov
exponent $-\sqrt{\beta}$ it is phase-space volume preserving as any
Hamiltonian motion. The exponent $\sqrt{\beta}$ relates to an
expanding eigenspace which is also the stable manifold; see
Fig.~\ref{f2}. A small coarse-graining will thus lead to an increasing
phase-space volume (and thus growing entropy) as demonstrated in
Fig.~\ref{f2}. It is expected that the rate of this increase
(i.e. entropy production) will be determined by
$\sqrt{\beta}\propto\sqrt{\Lambda}$. It was shown that under a weak
noise (which is equivalent to a specific coarse-graining) the positive
Lyapunov exponent defines the rate of entropy increase for the motion
in the inverted parabolic potential \cite{paz}. 

Recall that a coarse-graining (or alternatively an external noise) is
standardly needed for getting an arrow of time for a Hamiltonian
dynamics \cite{gal-or,lindblad,balian}. Hamiltonian systems displaying
the arrow of time are those, where a small coarse-graining leads to
entropy increase with a rate that only weakly depends on the
coarse-graining \cite{gal-or,lindblad,balian}.

Note that the $\propto\sqrt{\Lambda}$ scaling of the entropy
production is natural, since it indicates on the inapplicability of
the whole time-arrow scenario for $\Lambda<0$. Indeed, for
$\Lambda<0$, the cosmological constant amounts to an overall confining
potential and thus increases the stability of the motion.

However, this motion is not chaotic (not even ergodic). It generically
does become chaotic provided that its motion becomes bounded at some
scale; see \cite{fishman} for a concrete scenario related to the
inverted parabolic potential. It is conceivable that the accelerating
particle will meet other masses, and the resulting interaction will
achieve an effectively bounded phase-space. Similar and more complex
examples of hyperbolic motion in self-gravitating systems were studied
in \cite{gs,gk}. There is another argument showing that the motion in
the inverted-parabolic potential will change its dynamic regime: for a
very large $R$ the potential $\Phi(R)$ will assume very large absolute
values [cf.~(\ref{eq:1})], which violates the known applicability
condition $|\Phi|/c^2\ll 1$ of the non-relativistic description
\cite{nowak}.

It is known that in a Universe with a positive cosmological constant
$\Lambda>0$, the long-time evolution of the space-time will be
dominated by $\Lambda$ \cite{wald}. Locally (but not globally) this
Universe will look like empty, since the matter will escape through
the horizon \cite{wald}. Now for this Universe it was shown that the
generalized second law holds: the sum of the matter entropy and
geometric entropy (related to the horizon) does not decrease
\cite{gib,davies1,davies2}.  For very late times, where locally
(almost) no matter is present, the matter entropy will tend to zero,
while the geometric entropy saturates at a value $\propto 1/\Lambda$
\cite{gib,davies1,davies2}.  Our consideration concerns the matter
entropy and refers to the opposite limit of sufficiently early times,
where the matter is abundant, there are bound systems that perform
finite motion {\it etc}.

\section{Summary} 

Thus, the cosmological constant term induces hyperbolicity of motion
with the time asymmetry and dynamic entropy production on the
large-scale when $\Lambda$ is dominating. We provided a mechanism by
which at those scales $\Lambda$ facilitates the thermodynamic
arrow. We stress that this mechanism does not directly apply to early
universe. Indeed, our starting point is a closed (microcanonically
equilibrium) system that does not show an arrow of time before
perturbed externally. Such systems exists independently on initial
conditions of the early universe. 

We also note that the mechanism cannot (and should not) explain all
occurences of the thermodynamic arrow. However, note that even when
the dark energy (cosmological constant) does not dominate the mean
density (early universe or today's laboratory scale), it still
exists. To give an example: for a quantum system in a laboratory
$\Lambda$ (= vacuum energy) is not dominating, although it exists
(e.g. Casimir effect). Importantly, the dark energy can serve as an
ideal thermodynamic bath, since it does not get any back-reaction.

\end{document}